\documentclass[apjl]{emulateapj}

\usepackage{natbib}
\usepackage{amsmath,amssymb}
\usepackage{mathptmx}
\usepackage{graphicx}

\newcommand  \ergs     {\ifmmode {\rm erg\,s}^{-1} \else erg s$^{-1}$\fi}
\newcommand  \mbh      {\ifmmode M_{\rm BH} \else $M_{\rm BH}$\fi}
\newcommand  \cmii     {\ifmmode {\rm cm}^{-2} \else cm$^{-2}$\fi}
\newcommand  \cmiii     {\ifmmode {\rm cm}^{-3} \else cm$^{-3}$\fi}

\def\Hubble{\ifmmode {\rm km\,s}^{-1}\,{\rm Mpc}^{-1}
     \else km\,s$^{-1}$\,Mpc$^{-1}$\fi}

\def\Msun{\ifmmode M_{\odot} \else $M_{\odot}$\fi}
\def\Lsun{\ifmmode L_{\odot} \else $L_{\odot}$\fi}
\def\Zsun{\ifmmode Z_{\odot} \else $Z_{\odot}$\fi}

\def\qo{\ifmmode q_{0} \else $q_{0}$\fi}
\def\Ho{\ifmmode H_{0} \else $H_{0}$\fi}
\def\ho{\ifmmode h_{0} \else $h_{0}$\fi}
\def\qo{\ifmmode q_{0} \else $q_{0}$\fi}
\def\ao{\ifmmode a_{0} \else $a_{0}$\fi}
\def\to{\ifmmode t_{0} \else $t_{0}$\fi}
\def\omm{\ifmmode \Omega_{{\rm M}} \else $\Omega_{{\rm M}}$\fi}
\def\omlam{\ifmmode \Omega_{\Lambda} \else $\Omega_{\Lambda}$\fi}
\def\gtsim{\raisebox{-.5ex}{$\;\stackrel{>}{\sim}\;$}}

\def\mgii{\ifmmode {\rm Mg}{\textsc{ii}} \else Mg\,{\sc ii}\fi}
\newcommand \MgII {\ifmmode {\rm Mg}\,{\sc ii}\,\lambda2798 \else Mg\,{\sc ii}\,$\lambda2798$\fi}
\def\Hbeta{\ifmmode {\rm H}\beta \else H$\beta$\fi}
\def\civ{C\,{\sc iv}}

\newcommand{\lbol} {\ifmmode L_{\rm bol} \else $L_{\rm bol}$\fi}
\newcommand{\LHD} {\ifmmode L_{\rm HD} \else $L_{\rm HD}$\fi}
\newcommand{\lamLlam}{\ifmmode \lambda L_{\lambda} \else $\lambda L_{\lambda}$\fi}
\newcommand{\lledd}{\ifmmode L/L_{\rm Edd} \else $L/L_{\rm Edd}$\fi}

\newcommand{\fwmg}{\ifmmode {\rm FWHM}\left(\mgii\right) \else FWHM(\mgii)\fi}

\newcommand{\CFHD}{\ifmmode {\rm CF}_{\rm HD} \else ${\rm CF}_{\rm HD}$\fi}

\def  \mic         {$\mu$m}
\def  \MgII         {\ifmmode {\rm Mg}\,{\sc ii}\,\lambda2798
                  \else Mg\,{\sc ii}\,$\lambda2798$\fi}
\def  \mgii         {\ifmmode {\rm Mg}\,{\sc ii} \else Mg\,{\sc ii}\fi}

\def \spitzer      {{\it Spitzer}}

\def\Chisq{\ifmmode \chi^{2} \else $\chi^{2}$}
\begin{document}

\title{Hot Dust Clouds with Pure Graphite Composition around Type-I Active Galactic Nuclei}

\author{
Rivay Mor \altaffilmark{1}
\& Benny Trakhtenbrot \altaffilmark{1}
}

\altaffiltext{1}
{School of Physics and Astronomy and the Wise Observatory,
The Raymond and Beverly Sackler Faculty of Exact Sciences,
Tel-Aviv University, Tel-Aviv 69978, Israel}

\email{rivay@wise.tau.ac.il}

\begin{abstract}
We fitted the optical to mid-infrared (MIR) spectral energy distributions (SEDs) of $\sim$15000
type-I, $0.75<z<2$, active galactic nuclei (AGNs) in an attempt to constrain the properties
of the physical component  responsible for the rest-frame near-infrared (NIR) emission.
We combine optical spectra from the Sloan Digital Sky Survey (SDSS) and
MIR photometry from the preliminary data release of the Wide Infrared Survey Explorer (WISE). 
The sample spans a large range of AGN properties: luminosity, black hole mass, and accretion rate.
Our model has two components: a UV-optical continuum source and very hot, pure-graphite dust clouds.
We present the luminosity of the hot-dust component and its covering factor, for all sources,
and compare it with the intrinsic AGN properties.
We find that the hot-dust component is essential to explain the (rest) NIR emission in almost all AGNs in our sample, 
and that it is consistent with clouds containing pure-graphite grains and located between the dust-free broad line
region (BLR) and the ``standard'' torus.
The covering factor of this component has a relatively narrow distribution around a peak value of $\sim$0.13,
and it correlates with the AGN bolometric luminosity.
We suggest that there is no significant correlation with either black hole mass or normalized accretion rate.
The fraction of hot-dust-poor AGNs in our sample is $\sim15-20$\%, consistent
with previous studies. We do not find a dependence of this fraction on redshift or source luminosity.

\end{abstract}

\section{Introduction}
\label{sec_intro}

The unification scheme of active galactic nuclei (AGNs) requires an anisotropic obscuring structure
that surrounds the central accreting black hole \cite[e.g.,][]{Krolik1988,Antonucci1993}.
In this picture, the bulk of the radiation from the central engine is absorbed by the obscuring
structure (commonly referred to as the torus) and re-emitted mainly in mid-infrared (MIR) wavelengths.

A component of very hot dust at the innermost edge of the torus has been suggested
in the past \cite[e.g.,][]{Neugebauer+87, Barvainis87},
and in recent years is increasingly being supported by observations.
Reverberation measurements of nearby AGNs suggest that the near infrared (NIR) emission
in these sources is dominated by thermal radiation from hot dust very close to the central source \cite[few tens of light days; e.g., ][]{Minezaki+04,Suganuma+06}.
Other studies fitted the NIR-MIR spectral energy distributions (SEDs) of AGNs using a blackbody spectrum
to represent emission from hot dust in the inner region of the torus
(e.g., \citealt{Kishimoto+07, Riffel2009, Mor2009}, hereafter M09; \citealt{Deo2011}).
More recently \cite{Landt2011a} found similar results by fitting only the optical-NIR SED of 23 AGNs.
The modeled temperature of this component is found to be high, $\gtsim1200\,\rm{K}$,
regardless of the AGN luminosity and consistent with pure-graphite dust composition (M09).
Several studies have shown that the luminosity of the NIR excess emission correlates with 
that of the central engine \cite[e.g.,][and references therein]{Gallagher2007} with a slope close to unity. 
However, it is yet unclear whether this hot-dust component is related to other 
AGN properties such as its mass (\mbh) or normalized accretion rate (\lledd).

Although hot dust seems to be a common feature of AGNs, several recent studies have
suggested that a certain fraction of the AGN population lacks such a component.
\cite{Jiang2010} found two z$\simeq$6 QSOs without any detectable emission from hot dust
(using \spitzer\ MIR photometry). These were dubbed ``hot-dust free'' QSOs.
\cite{Hao2010} and more recently \cite{Hao2011} found a sizable amount of type-I AGNs with unusually weak NIR emission in
several large samples of type-I AGNs. These were dubbed ``hot-dust-poor'' (HDP; hereafter we adopt this notation) AGNs.
These authors found that the fraction of HDP AGNs increases with redshift from 6\% at low redshift ($z<2$) to 20\% at higher redshift ($2<z<3.5$).
We note that all the studies mentioned here are limited to relatively small samples, and use highly simplified emission models.

In this letter we use the recently published preliminary data release of the Wide Infrared Survey Explorer
\cite[WISE;][]{Wright2010}, together with the seventh data release of the Sloan Digital Sky Survey \cite[SDSS/DR7;][]{Abazajian2009} 
to construct $\sim15000$ UV to NIR SEDs of type-I AGNs.
We apply spectral decomposition using novel models to deduce the
properties of the hot dust, and test these properties against different intrinsic AGN properties.


\section{Sample Selection and Spectral Decomposition}
\label{sec_sample}
%

We aim to study a large and uniformly selected sample of type-I AGNs, for which \lbol, \mbh\ and \lledd\ can
be reliably measured, and the WISE bands (at $\sim3.4,\,4.6$, \& 11.6 \mic) cover the rest-frame
wavelength range of 1 to 5 \mic, where the emission originating from the hot dust is expected to peak.
At longer wavelengths (i.e., those covered by the $\sim22$ \mic\ WISE band),
the emission from the ``standard'' torus becomes significant (see Fig.~4 in M09).
Thus, we select from the SDSS only high confidence ``QSO'' sources (i.e. \texttt{zconf}$>0.7$) at $0.75<z<2$.
This query returned 20231 sources within the area covered in the WISE preliminary DR.
Cross-matching this sample with the WISE catalog, using a 3\arcsec\ search radius, yielded 19116 matched sources. 
The remaining 1115 sources ($\sim5.5\%$) probably lie below the flux limit defined for the WISE preliminary DR, of $\sim0.08\,\rm{mJy}$ 
in the 3.4 \mic\ band. 
These sources have a similar range of \lbol\ to that of the whole parent sample, and are further discussed in \S\ref{sec_results}.
Thus, our analysis of the WISE data provides useful information regarding $\sim94.5\%$ of optically selected, $0.75<z<2$, type-I AGN (see more criteria below).

We further filtered the resultant cross-matched catalog to include only the 17920 sources which have $S/N>3$ in each of the
$\sim$3.4, 4.6 \& 11.6 \mic\ WISE bands. Our analysis ignores the $\sim22$ \mic\ band.
The WISE cataloged magnitudes and associated uncertainties were translated to $f_\nu$ through the published (``iso'') WISE zero points.
\footnote{See: http://wise2.ipac.caltech.edu/docs/release/prelim/expsup/figures/\\sec4\_3gt4.gif}
The SDSS spectra of all these QSOs were modeled by a comprehensive procedure that is aimed at fitting
the broad \MgII\ emission line and the adjacent emission complexes.
The procedure is identical to that used in \cite{Trakhtenbrot2011}, and similar to other studies \cite[e.g.,][]{Shen2008, Fine2008}.
Most importantly, it allows a robust determination of FWHM(\mgii) and of the monochromatic luminosity at
(rest-frame) 3000\AA\ ($\lambda\,L_{\lambda}[3000$\AA$]$, hereafter $L_{3000}$).
Further 1571 sources, for which the \mgii\ fitting procedure failed to provide high-confidence results, were excluded.
We also excluded 421 sources with $\log\left(L_{3000}/\ergs\right)<45$, 
which might contain some host-galaxy contribution to their rest-frame NIR SED, and often have SDSS spectra with low S/N.
Finally, our sample consists of 15928 QSOs with $45<\log\left(L_{3000}/\ergs\right)<47.25$, 
which correspond to about 78\% of all SDSS selected $0.75<z<2$, type-I AGN covered by WISE observations.
From these measurements, we deduce \mbh\ using the \cite{McLure2004a} prescription.
\lbol\ is calculated from $L_{3000}$ by applying a luminosity-dependant bolometric correction,
which will be discussed in full detail in a forthcoming publication. 
Here we only note that these correction factors were independently calibrated using a large sample
of $0.5<z<0.75$ QSOs, for which \lbol\ could be reliably derived by the \cite{Marconi2004} prescription.
The bolometric corrections applied to the SDSS/WISE sample range between about 3, for the most luminous sources, and 4.2, for the faintest.
These values are lower, by a factor of $\sim1.5-1.7$, than those of \cite{Richards2006} or \cite{Elvis1994}, since they avoid the 
double-counting of the re-processed IR emission \cite[see the detailed discussion in][]{Marconi2004}.
The resultant range in $\log \left(\lbol/\ergs\right)$ is $45.6-47.7$.
We estimated the accretion rate as $\lledd=\lbol/\left(1.5\times10^{38}\,\mbh\right)$.
%


The SDSS/WISE SED of each source in our sample is fitted to constrain the properties of the hot-dust component
that dominates the NIR wavelength range.
To focus on continuum emission, every SDSS spectrum was sampled at several continuum bands, around the (rest frame) wavelengths: 
$1450-1475,\,2150-2200,\,3030-3100,\,4150-4250,\,5080-5120,\,5600-5750,\,\&\,6100-6250$ \AA\ \cite[e.g.,][]{VandenBerk+01}.
Our models include two different components: a source of UV-optical continuum emission (e.g., an accretion disk)
and very hot-dust clouds with pure graphite composition.

\begin{figure}
\includegraphics[width=0.4\textwidth]{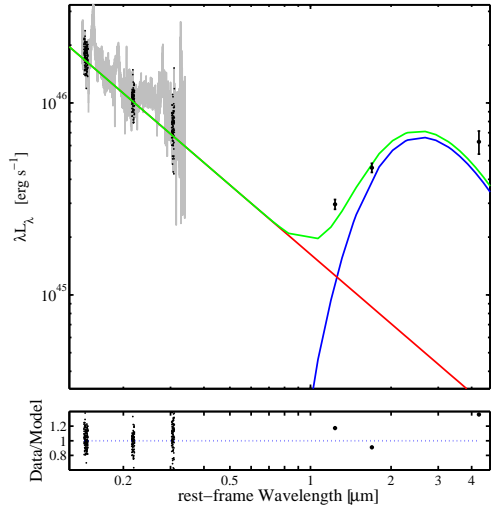}
\caption{Best fit to the SED of SDSS J134154.22+005948.6 ($z=1.715$) using a combination
of power-law (red line) and hot, pure-graphite dust (blue line) components.
The original SDSS spectrum is shown in gray and the continuum bands selected for the 
fitting procedure are highlighted in black. The best fit model is shown in red.
The quality of the fit is demonstrated by the ratio between the data and the model (bottom).}
\label{fig:fit_example}
\end{figure}

To model the central-source continuum emission we assume a power-law function, $L_{\nu}\propto\nu^{\alpha}$.
This component has two free parameters, the power-law index and the normalization factor.
The power-law index, $\alpha$, is changed in steps of 0.1 between -1.5 to 1.
The second component represents a collection of dusty clouds of gas with a pure-graphite grain composition.
The SED of this component is taken from the models of \cite{Mor2011b}, which provide a full discussion of all model assumptions and parameters.
In short, this study adopts the ``cloud model'' of the broad-line region (BLR) and considers the graphite-containing gas to be the extension of the BLR.
The model assumes gas composition of 2 \Zsun, standard ISM-type dust depletion, gas density in the range $\log\left(n/\cmiii\right)=9.3-9.8$ and 
column density in the range $\log\left(N_{\rm H}/\cmii\right)=22.3-22.7$ at the graphite sublimation radius.
While the local emission of the graphite dust depends only on the grain properties, the dust temperature
inside the cloud varies by a large factor because of the local grain opacity.
Thus a single cloud spectrum appears as a combination of many modified blackbodies. 
The results of the calculations provide a grid of SEDs that are used in the fitting procedure.
For this component there are two free fitting parameters, the distance to the cloud (given a source luminosity)
and a normalization factor which determines its luminosity (hereafter \LHD).
For example, the range of distances for an AGN with $\log\left(\lbol/\ergs\right)=46.9$ is 1.3 to 3.5 pc.
The fitting procedure uses a standard \Chisq\ minimization to determine the best fit
combination of power-law and hot-dust model.
Figure~\ref{fig:fit_example} demonstrates the fitting procedure for a representative case.
The hot pure-graphite dust component dominates the SED between $\sim$2 and 5 \mic.
The main caveat of the fitting procedure is the (unknown) contribution to the SED from 
cooler dust in the torus.
As shown in \cite{Mor2011b}, neglecting this component may lead to an overestimation of \LHD\ by merely $\sim10-20$\%.
In the following analysis, we do not correct for this systematic uncertainty.

The chosen redshift range, and thus typical AGN luminosities, securely omits any host galaxy dominated SEDs from our sample.
To minimize the effect of the (unknown) host-related extinction, we further limit our sample
to include only sources for which the best-fit slope of the power-law component satisfies $\alpha>-0.8$.
This choice is comparable to those made in other studies of the optical to NIR SED of AGNs.
For example, the \cite{VandenBerk+01} composite has $\alpha\simeq-0.44$ over the relevant wavelength range, 
while \cite{Hao2010} show that $\alpha$ ranges between about -0.8 and 0.6, based on the \cite{Elvis1994} SEDs.
In any case, omitting this criterion adds only a small fraction of sources and does not significantly change our main results.
A more comprehensive analysis of the extinction is beyond the scope of this \textit{Letter}.
The removal of these sources leaves 15077 sources that will be used in the following analysis.


\section{Results and Discussion}
\label{sec_results}
\begin{figure}
\includegraphics[width=0.45\textwidth]{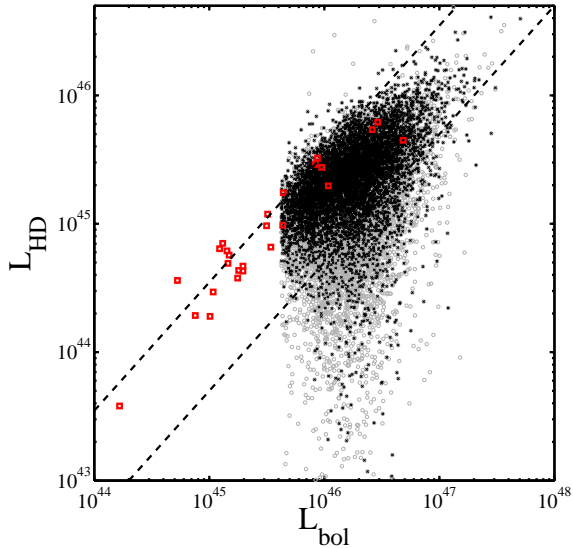}
\caption{Hot-dust luminosity vs. AGN bolometric luminosity.
Results from the complete WISE sample studied in this work are shown as gray symbols.
Black symbols highlight the sources with $\alpha>-0.4$.
The QUEST sample of PG QSOs are shown in red. The dashed lines represent $\LHD\propto\lbol$ 
(i.e. constant \CFHD), with normalization of 0.05 and 0.35.}
\label{fig:LHD_Lbol}
\end{figure}

The most important parameter of the hot-dust component, determined by the fitting procedure, is \LHD.
Figure~\ref{fig:LHD_Lbol} shows a prominent correlation between \LHD\ and \lbol.
We also highlight a sub-sample of sources with extremely blue continua, which are defined as those with $\alpha>-0.4$, and
correspond to the bluest third of the sample \cite[following][]{Gallagher2007}. 
These sources are presumably less affected by extinction.
The distribution of these blue sources in the $\LHD-\lbol$ plane is similar to that of the entire sample, 
suggesting that host extinction plays only a minor role in our analysis.
Results for the QUEST sample (PG QSOs studied in M09; red squares in Fig.~\ref{fig:LHD_Lbol}) were calculated by 
employing the M09 procedure (using the pure-graphite models) to the $\sim$2 to 35 \mic\ SEDs.

\begin{figure}
\includegraphics[width=0.4\textwidth]{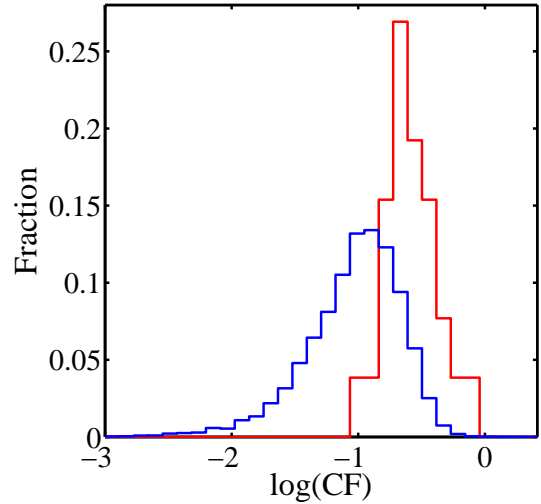}
\caption{Distribution of the hot-dust covering factor.
Values of the current study (blue) are generally smaller than those found for the
QUEST sample (red) and may be the result of the different luminosity range of
the two samples (see Figure~\ref{fig:CF_Lbol}).
The \CFHD\ distribution is asymmetric with a clear excess of low \CFHD\ sources.
The peak of the distribution is at $\sim$0.13 and the median value is 0.1.
Sources with \CFHD\ below a certain luminosity-dependant value may represent the so-called hot-dust-poor or dust-free AGNs.}
\label{fig:hist_CF}
\end{figure}

Assuming that the rest-frame NIR SED is due to reprocessed AGN radiation by the hot-dust clouds,
we can deduce the covering factor (\CFHD) of the central source by these clouds, defined by $\CFHD=\LHD/\lbol$  \cite[][M09]{Maiolino2007a}.
%
Fig.~\ref{fig:LHD_Lbol} clearly indicates that \CFHD\ of the entire population spans a relatively limited range.
Indeed, the distribution of \CFHD\ (Figure~\ref{fig:hist_CF}) appears to be fairly concentrated
around a peak value of 0.13 (median value is 0.1 with a standard deviation of 0.4 dex).
This value is smaller than the one found for the lower luminosity, lower redshift, QUEST sample ($\sim$0.23).
The distribution of \CFHD\ is asymmetric with a clear excess of sources with low \CFHD. 
These sources may represent a different subset of the population with a very small or no hot-dust component.
We note that for a small fraction of the sample ($\sim1.7\%$, 258 sources) the fit does not require a hot-dust 
component thus resulting in $\LHD=0$ and $\CFHD=0$. 
We add these sources to the sub-group of HDP AGNs discussed below.

\begin{figure*}[ht]
\includegraphics[width=0.95\textwidth]{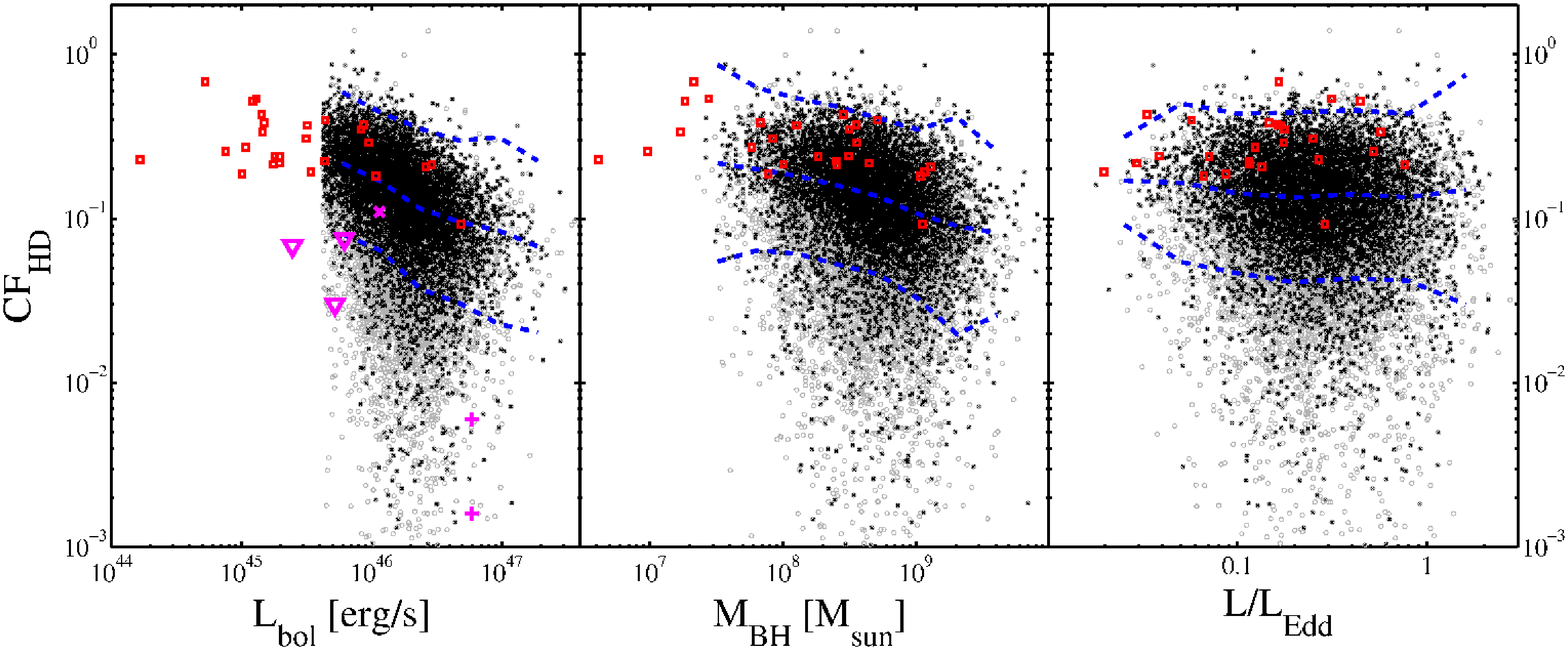}
\caption{Hot-dust covering factor vs. AGN properties.
Symbols as in Fig.~\ref{fig:LHD_Lbol}.
Left: \CFHD\ decreases with increasing \lbol. 
Magenta markers represent the mean SED of \cite{Elvis1994}, scaled to $\log\left(\lbol/\ergs\right)=46$ (cross);
the three examples of hot-dust-poor objects of \citet[][triangles]{Hao2010}
and the two $z\simeq6$ dust-free QSOs of \citet[][plus signs]{Jiang2010}.
The correlations with \mbh\ and \lledd\ (middle and right panels) are much less significant.
In all panels, dashed black lines represent the peak of the \CFHD\ distribution and
the 3$\sigma$-equivalent range - see text}
\label{fig:CF_Lbol}
\end{figure*}

Fig.~\ref{fig:CF_Lbol} presents a clear anti-correlation between \CFHD\ and \lbol.
This trend is further confirmed by both Pearson's and Spearman's rank correlation tests (p value $\ll0.01$ for both tests).
In particular, the typical \CFHD\ decreases from $\sim0.2$ for sources with
$\log\left(\lbol/\ergs\right)\simeq45.6$ to $\sim0.09$ for sources with $\log\left(\lbol/\ergs\right)\simeq46.6$.
We verified that a similar relation is recovered even if one calculates \lbol\ with the commonly
used, uniform (and overestimated; see \S\ref{sec_sample}) bolometric correction of 5.15 or 5.62 \cite[e.g.,][]{Elvis1994,Richards2006}.
Several earlier studies suggested a similar trend \cite[e.g.,][]{Wang2005, Maiolino2007a, Treister2008},
however these were based on the total MIR emission which translates to the covering factor of the entire dusty 
structure, and not just the hot dust.
\cite{Gallagher2007} suggest that this $\CFHD-\lbol$ anti-correlation may be a manifestation of dust extinction. 
We therefore verified that this trend persists for the sub-sample of blue sources, defined above.
We further checked whether the \CFHD-\lbol\ relation is due to the higher abundance of more luminous
AGNs towards higher redshifts, using a sub-sample of 3837 sources which reside in a narrow
luminosity range (0.2 dex) around $\log\left(\lbol/\ergs\right)\simeq46.3$.
Although this sub-sample spans almost the entire redshift range of the original sample, no correlation is found between \CFHD\ and redshift.

The physical mechanism responsible for the decrease of covering factor with \lbol\ is still undetermined.
One possibility is a ``receding torus'' scenario \cite[][]{Lawrence1991}, where higher luminosity implies larger
dust sublimation distance, and hence an obscuring structure that is located farther away from the center.
In this scenario, however, the geometry of the hot-dust clouds must be toroidal and have a constant height.
Another possibility is that the \CFHD-\lbol\ anti-correlation is analogous to the anti-correlation found 
between the equivalent width of different BLR lines (\Hbeta\ and \civ) and AGN luminosity \cite[e.g.,][]{Netzer2004, Baskin2005}.

The correlations between \CFHD\ and \mbh\ and \lledd\ (Fig.~\ref{fig:CF_Lbol}, middle and right panels) are much less significant. 
We interpret these as the result of the marginal dependence of both \mbh\ and \lledd\ on the source luminosity.
This is further confirmed by the fact that these marginal correlations
disappear completely in the $\log\left(\lbol/\ergs\right)\simeq46.3$ sub-sample mentioned above.
However, the large scatter makes it difficult to securely determine which correlation is more fundamental.
We suggest that the apparent \CFHD\ is sensitive to the immediate irradiating flux from the central source,
but does not depend on the evolutionary stage of the accreting super-massive black hole (SMBH).


For a certain fraction of the sources, the hot-dust a component is much less prominent. 
First, as noted above, $\sim1.7\%$ of the sources have $\LHD=0$ and thus $\CFHD=0$. These sources spread the entire range of \lbol.
Second, there is a non negligible population of sources with significantly
lower \CFHD\ values than the main locus of sources (Fig.~\ref{fig:CF_Lbol}).
There is no corresponding population of high-\CFHD\ outliers.
Fig.~\ref{fig:CF_Lbol} illustrates this using the boundaries of the 99th percentiles of the \CFHD\ distribution.
The percentiles are calculated by assuming that the \CFHD\ distribution should be symmetrical around
the peak and mirroring the high-\CFHD\ side of the distribution. This is done in each (0.2 dex) luminosity bin separately.
We suggest that all the points which lie below the lower dashed line in Fig.~\ref{fig:CF_Lbol}
can be regarded as HDP AGNs. Thus, we are able to provide an HDP criterion which depends on \lbol.
For example, for $\log\left(\lbol/\ergs\right)=46$ all sources with $\CFHD\lesssim0.06$ may be considered as HDP AGNs.

The total fraction of HDP AGNs in our sample is $\sim16.2\%$ (2439 sources out of 15077, including the 258 sources with $\CFHD=0$).
This number is in rough agreement with that reported by \cite{Hao2010} and \cite{Hao2011}. 
We do not find a significant dependence of this fraction with either luminosity or redshift. 
We stress that using a HDP criterion that is independent of AGN luminosity can result in false correlations
of the fraction of HDP AGNs with either luminosity or redshift. 
The latter may be a consequence of selection biases of high luminosity sources towards higher redshifts.
As mentioned in \S\ref{sec_sample}, 1115 sources ($\sim5.5\%$ of the total SDSS detections)
probably lie below the flux limit defined for the preliminary DR of WISE.
These sources may also represent HDP AGNs, thus should be added to the total fraction mentioned above ($\sim16.2\%$).

As mentioned in \S\ref{sec_intro}, \cite{Jiang2010} suggested that a significant fraction of $z\simeq6$ AGNs are hot-dust free,
i.e. lacking any emission from a hot, dusty component. Moreover, these authors claim that such systems are not observed at lower redshifts.
Taken at face value, our $\CFHD=0$ sources can be regarded as the $z<2$ counterparts of the $z\simeq6$
hot-dust free AGNs. These are indeed rare at low redshifts.
However, the hot-dust free AGNs of \cite{Jiang2010} have only upper limits in the (rest-frame) IR bands, 
which place them in our HDP AGN regime (see Fig.~\ref{fig:CF_Lbol}).
Thus, our sample might offer many more $z<2$ sources that correspond to the hot-dust free notion of \cite{Jiang2010},
from both the WISE-undetected sub-sample ($\sim5.5\%$) and the HDP AGN sub-sample ($\sim16.2\%$).
We also note that variability between the SDSS and WISE observational epochs would not effect the 
estimation of the fraction of HDP AGNs for large samples. It may, however, account for some of the scatter
observed in the symmetric part of the \CFHD\ distribution.

We have demonstrated that the rest-frame NIR emission of the vast majority ($\sim80$\%) of type-I AGNs
in our unprecedentedly large sample can be explained by emission from hot, pure-graphite dust clouds. 
These clouds reside at distances greater by a factor of about $3-10$ than the BLR clouds, 
and have a typical \CFHD\ of $\sim0.13$.
The lack of a significant correlation between \CFHD\ and \mbh\  
stands in contrast to the result of \cite{Jiang2010}, who suggested 
that such a relation only exists at very high redshifts ($z\simeq6$), where dusty structures around AGNs 
were not yet fully developed.
Interestingly, we do not find any significant correlation between \CFHD\ and either \lledd\ or redshift,
despite the fact that our sample effectively covers a wide range in these properties ($0.05\lesssim\lledd$ and $0.75<z<2$).
We suggest that previous reports of a decrease in \CFHD, and an increase in the fraction of HDP AGNs, with increasing redshift, 
may be due to a combination of the selection of high-luminosity sources at high redshifts and the \CFHD-\lbol\ anti-correlation
reported here. 
However, the small overlap in \lbol\ between the low-z QUEST sample and the SDSS/WISE sample, 
and the limited range of \lbol\ for SDSS/WISE sources, at a chosen redshift, inhibits us from testing this issue further.

To fully understand the \textit{evolution} of hot dusty structures around accreting SMBHs, all these trends should 
be tested in large samples of high redshift type-I quasars, for which \lbol, \mbh\ and \lledd\ were reliably measured
\cite[e.g.,][]{Netzer2007a, Willott2010, Trakhtenbrot2011}.
As demonstrated in this \textit{Letter}, WISE data provide the optimal way to preform such studies, 
owing to its wide areal coverage and relatively deep flux limit. In addition, WISE covers the wavelength range where
hot-dust emission dominates the SED of AGNs, thus providing more direct probes of this emission at high-redshift sources
than NIR data.


\begin{acknowledgements}
We thank Hagai Netzer for providing the radiative models used in this work and for many helpful comments.
Funding for this work has been provided by the Israel Science Foundation grant 364/07 and the 
DFG via German-Israeli Cooperation grant STE1869/1-1.GE625/15-1.
This publication makes use of data products from the Wide-field Infrared Survey Explorer,
which is a joint project of the University of California, Los Angeles, and the
Jet Propulsion Laboratory/California Institute of Technology, funded by the National
Aeronautics and Space Administration.
This study makes use of data from the SDSS (see: http://www.sdss.org/collaboration/credits.html).
\end{acknowledgements}




\end{document}